\documentclass[prl,twocolumn,showpacs,superscriptaddress,aps]{revtex4-1}
\usepackage{graphicx,amssymb,amsmath,color,psfrag}
\usepackage[colorlinks=true,citecolor=blue,linkcolor=blue,urlcolor=blue]{hyperref}
\begin{document}
\definecolor{darkgreen}{rgb}{0,0.5,0}
\newcommand{\be}{\begin{equation}}
\newcommand{\ee}{\end{equation}}
\newcommand{\jav}[1]{#1}

\title{Detecting equilibrium and dynamical quantum phase transitions \jav{in Ising chains} via out-of-time-ordered correlators}

\author{Markus Heyl}
\affiliation{Max-Planck-Institut f\"ur Physik komplexer Systeme, 01187 Dresden, Germany}
\author{Frank Pollmann}
\affiliation{Department of Physics, Technical University of Munich, 85748 Garching, Germany}
\author{Bal\'azs D\'ora}
\email{dora@eik.bme.hu}
\affiliation{Department of Theoretical Physics and MTA-BME Lend\"{u}let Spintronics 
Research Group (PROSPIN), Budapest University of Technology and Economics, 1521 Budapest, Hungary}

\date{\today}

\begin{abstract}
Out-of-time-ordered  (OTO) correlators have developed into a central concept quantifying quantum information transport, information scrambling and quantum chaos. 
In this work we show that such OTO correlator can also be used to dynamically detect equilibrium as well as nonequilibrium phase transitions \jav{in Ising chains}.
We study OTO correlators of an order parameter both in equilibrium and after a quantum quench \jav{for different variants of transverse-field Ising models in one dimension}, including the integrable one as well as non-integrable and long-range extensions.
\jav{We find for all the studied models that the OTO correlator in ground states detects the quantum phase transition.}
\jav{After a quantum quench from a fully polarized state we observe numerically for the short-range models that the asymptotic long-time value of the OTO correlator signals still the equilibrium critical points and ordered phases. For the long-range extension, the OTO correlator instead determines a dynamical quantum phase transition in the model.}
We discuss how our findings can be observed in current experiments of trapped ions or Rydberg atoms.
\end{abstract}

%\pacs{71.10.Pm,03.67.Mn}

\maketitle

\paragraph{Introduction.} Today, synthetic quantum materials such as ultra-cold atoms or trapped ions can experimentally access quantum dynamics governed by purely unitary evolution with a negligible coupling to an environment on the relevant time scales~\cite{BlochDalibardZwerger_RMP08, Bloch2012qs, Blatt2012, Georgescu2014}.
This has lead to the observation of quantum dynamical phenomena such as many-body localization~\cite{Schreiber2015oo, Smith2016, Bordia2016, Choi2016}, particle-antiparticle production in the Schwinger model~\cite{Martinez2016}, dynamical quantum phase transitions~\cite{Flaeschner2016, Jurcevic2017, Zhang20172}, or discrete time crystals~\cite{Zhang2017, Choi2017}.
In many of these phenomena the propagation of quantum information plays a central role such as for the celebrated logarithmic entanglement growth in many-body 
localized systems~\cite{Znidaric2008,Bardarson2012}.
For the quantum formation transport captured by quantum correlations Lieb-Robinson bounds~\cite{lieb1972,Hastings2006,Nachtergaele2006} give fundamental constraints, which can be lifted only for long-ranged interacting systems~\cite{Hauke2013so} as 
demonstrated also experimentally~\cite{Richerme2014,Jurcevic2014}.
Recently, it has been realized that out-of-time-ordered (OTO) correlation functions can capture information propagation beyond quantum correlation spreading~\cite{junli, garttner, nyao, swingle, zhugrover, campisi, aleiner, bohrdt,fan2017,huang2017}.
In particular, such OTO correlators can diagnose quantum chaos in terms of information scrambling via an exponential growth bounded by a thermal Lyapunov exponent~\cite{maldacena2016}.
In this work we show, that OTO correlators can also be used to dynamically detect both equilibrium and dynamical quantum phases and the associated quantum critical points.
Specifically, we study the OTO correlator dynamics in equilibrium states and after quantum quenches \jav{in transverse field Ising chains with and without long range couplings.}
When choosing as the operator in the OTO correlator the order parameter of the underlying transition, 
we find that the long-time limit serves as a diagnostic tool to detect phases and transitions:
In the symmetry-broken phase the OTO correlator is nonzero and vanishes upon approaching the critical point remaining zero in the full paramagnetic phase.
In this way, one can possibly detect quantum criticality \jav{ in one-dimensional systems, that do not exhibit symmetry-breaking at nonzero temperatures, 
without preparing the system in the actual ground state, providing a purely dynamical signature of equilibrium quantum phases.}
We demonstrate our findings for both an integrable as well as nonintegrable version of the one-dimensional transverse-field Ising chain.
\jav{As a system with a finite-temperature phase transition we additionally study a long-range transverse-field Ising model where we find that the OTO correlator in equilibrium states correctly captures the equilibrium phases. In the dynamics after a quantum quench we find that the OTO correlator probes instead a dynamical quantum phase transition  of genuine nonequilibrium nature in the system's long-time steady state~\cite{Sciolla2011, Zunkovic2016, Zhang20172}.}
We also discuss how our results can be observed in current experiments in systems of trapped ions or Rydberg atoms.

OTO correlation functions~\cite{larkin} have been identified as quantities providing insights into quantum chaos and information scrambling~\cite{maldacena2016}.
The OTO commutator is defined as
\begin{gather}
C(t)=-\left\langle \left[V^{},W(t)\right]^2\right\rangle\geq 0,
\label{ct}
\end{gather}
where $V$ and $W$ are usually chosen as local Hermitian operators and $W(t)=\exp(iHt)W\exp(-iHt)$ with $H$ the system Hamiltonian.
The OTO commutator contains terms of the form $\mathcal{F}(t)=\langle W(t)VW(t)V)\rangle$, coined  OTO correlator due to its unconventional temporal structure.
These quantities probe spread of quantum information beyond quantum correlations, in particular signaling the presence
of quantum chaos, with a growth bounded by a thermal Lyapunov
exponent~\cite{maldacena2016}.
Recently, much effort, including experiments~\cite{junli,garttner,nyao,swingle,zhugrover,campisi,aleiner,bohrdt},
has been devoted to studying its behaviour, with peculiar links to the physics of black holes and
random matrix theory~\cite{maldacena2016,cotler}. Additionally, a simple 'mesoscopic' Sachdev-Ye-Kitaev model~\cite{sachdevye,kitaev17,maldacenaprd,larkin,fritzpatrick}
captures many interesting phenomena, including a maximal Lyapunov exponent and entropy characteristic to black holes.

We investigate numerically such OTO correlators in a variety of one-dimensional, exhibiting equilibrium and dynamical quantum phase transitions of different kinds.
We choose as operators $V=W=\mathcal{M}$ the order parameters $\mathcal{M}$ of the respective transitions in the considered models which in all of the considered cases is a magnetization
\begin{gather}
\mathcal{M} =\left\{ \begin{array}{c}
\sigma_n^z \textmd{ for short-range models}\, ,\\
S^z  \textmd{ for \jav{the} collective spin model} \, ,
\end{array}
\right.
\label{defM}
\end{gather}
where $\sigma_n^z$ are Pauli matrices and $n=1,\dots,N$ with $N$ is the total number of lattice sites of the system and $S^z = N^{-1} \sum_{n=1}^N \sigma_n^z$ is the total spin operator.
Concretely, we study the dynamics of OTO correlators of the form
\begin{gather}
\mathcal{F}(t)=\langle \mathcal{M}(t)\mathcal{M}\mathcal{M}(t)\mathcal{M}\rangle
\label{czzzz}
\end{gather}
with the expectation value $\langle \dots \rangle = \langle \psi_0 | \dots | \psi_0\rangle$.
For $|\psi_0\rangle$ we choose two different states.
First, we take the respective ground state of the model at the given parameter set in order to probe the equilibrium phase diagram.
Second, for the study of the nonequilibrium dynamics we choose a fully polarized state $|\psi_0\rangle = |\uparrow\uparrow\uparrow\dots\rangle$, which on the one hand can be prepared in experiments of trapped ions or Rydberg atoms with high fidelity~\cite{Lanyon2011,Schauss2012,Jurcevic2017,Bernien2017,Guardado2017, Zhang2017, Zhang20172} and on the other hand is well suited to study dynamical quantum phase transitions (DQPTs) in nonequilibrium time evolution with the considered Ising models~\cite{Sciolla2011, heyl, Zunkovic2016, Jurcevic2017, Zhang20172}. % motivated by recent experiments in trapped ions and Rydberg atoms for two reasons
%\footnote{From the numerical point of view, this state is represented perfectly in finite size systems.}.
%
%First, this state can be prepared with high fidelity~\cite{Lanyon2011,Schauss2012,Jurcevic2017,Bernien2017,Guardado2017}.
%
%Second, for that case $\mathcal{F}(t)$ as defined in Eq.~(\ref{czzzz}) describes the OTO dynamics after a quantum quench which, in principle, can be accessed experimentally.
%
Details on how to access experimentally our theoretical predictions we give in the concluding discussion.

\paragraph{Transverse field Ising chain.} Let us start with the paradigmatic model for quantum phase transitions, the 1D transverse 
field Ising (TFI) chain~\cite{sachdev}, whose dynamics has been realized in recent experiments of Rydberg atoms when interactions beyond nearest neighbors can be neglected on the relevant time scales~\cite{Henning2016,Bernien2017,Guardado2017}.
Its Hamiltonian with periodic boundary condition reads as
\begin{gather}
H=-J\sum_{n=1}^N\sigma^z_n\sigma^z_{n+1}+g\sum_{n=1}^N\sigma^x_n,
\label{tfi}
\end{gather}
where the $\sigma^i_n$'s are Pauli matrices and $\sigma^i_{N+1}=\sigma^i_{1}$ with $i=x,y,z$. This model hosts an equilibrium quantum phase transition (QPT) at $g=J$ separating a 
paramagnetic phase for $g>J$ from a symmetry-broken phase with nonzero magnetization along the $\sigma^z$ direction for $g<J$~\cite{sachdev}.
For quantum quenches the system exhibits the appearance of DQPTs with nonanalytic behavior during quantum real-time dynamics whenever the quench crosses the underlying equilibrium QPT~\cite{heyl,heylreview}.

The ordered phase can be detected in equilibrium from dynamics by the autocorrelation function $\langle\sigma_n^z(t)\sigma_n^z\rangle$ which takes a nonzero/vanishing value 
in the long-time limit 
in the ferromagnetic/paramagnetic phase, respectively~\cite{sachdev}. However, in the case of a quantum quench, it becomes fully featureless since it vanishes for 
long times~\cite{essler,EPAPS}, 
irrespective of the Hamiltonian parameters.
It would nevertheless be advisable to detect both the QPT or DQPT from a dynamical measurement because these are naturally accessible experimentally in quantum simulators.
Since the autocorrelation function does not fulfill this job, it looks natural to try its second moment, i.e. $\langle(\sigma_n^z(t)\sigma_n^z)^2\rangle$, which is nothing but the OTO correlator discussed before. 

\jav{The model can be mapped onto free fermions such that many correlation functions can be calculated in a simple analytical manner, except for the order parameter $\sigma^z_n$~\cite{essler}.  Therefore,}
we calculate the OTO correlator using numerical methods, such as time evolving block decimation (TEBD) \cite{Vidal2003a} and exact diagonalization (ED).
Still, it can be evaluated exactly analytically in certain limiting cases:
for $g=0$, it takes its maximal value 1, while vanishes 
in the $J=0$ limit\footnote{For $g=0$, i.e. deep in the symmetry broken phase, the OTO correlator  
takes on its maximal value. The operators $\sigma_n^z$ are constants of motion since the Hamiltonian for $g=0$ contains only $\sigma^z$, 
and already the relation $\sigma^z(t)\sigma^z\sigma^z(t)\sigma^z=(\sigma^z)^4=1$ holds at any given site as an operator identity.
In the opposite, $J=0$ limit, the long time average of the same correlator vanishes, irrespective 
of the ensemble over which the expectation value is taken. This follows from the fact that $\sigma^z(t)\sigma^z\sigma^z(t)\sigma^z=\cos(4tg)+i\sigma_x\sin(4tg)$,
and upon taking the long time average, both terms vanish identically.}.
In between these two limits, the OTO correlator is expected to interpolate.
Whether the transition occurs at the critical point or at some other location is an intriguing question that we investigate in the following.

\begin{figure}[h!]
%\psfrag{x}[t][][1][0]{$tJ$}
%\psfrag{y}[b][][1][0]{$\mathcal{F}(t)$}
%\psfrag{t1}[][][1][0]{equilibrium, $g/J$=0.5}
%\psfrag{t2}[][][1][0]{equilibrium, $g/J$=1.5}
%\psfrag{t3}[][][1][0]{quench, $g/J$=0.5}
%\psfrag{t4}[][][1][0]{quench, $g/J$=1.5}
\includegraphics[width=8cm]{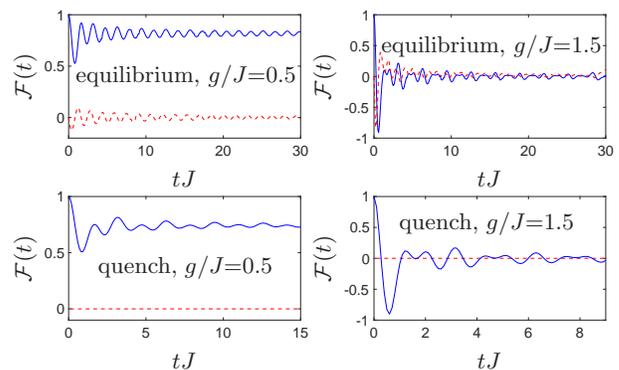}
\caption{The representative time evolution of the real (blue solid line) and imaginary (red dashed line) parts of OTO correlator are 
shown in equilibrium in the ordered phase with $g/J=0.5$ and
disordered phase with  $g/J$=1.5 from TEBD with $N=60$.
The bottom row visualizes the time evolution following a quench quench from the fully polarized  state from ED with $N=20$, before finite size effects appear.
\label{timedepotoc}}
\end{figure}

This we study numerically on finite systems, consisting of up to  $N=60$ spins in equilibrium for 
TEBD and up to 22 spins for ED. The time dependence of Eq. \eqref{czzzz} is shown in Fig. \ref{timedepotoc} for several representative parameter sets both in equilibrium and
after a quantum quench, in the ordered and disordered phase. 
While the real part steady state value of $\mathcal{F}(t)$ depends on whether the time evolving Hamiltonian is in the ordered or disordered region, its imaginary part vanishes identically in the steady state.
After a quantum quench we find that $\mathrm{Im} \, \mathcal{F}(t)$=0 such that we focus on the \emph{real} part of the OTO correlator $\mathcal{F}_R(t) = \mathrm{Re} \, \mathcal{F}(t)$ in the following.
$\mathcal{F}_R(t)$ starts from $\mathcal{F}_R(t=0)=1$ due to the operator identity $(\sigma_n^z)^2=1$, and reaches rather quickly a time-independent steady state value before finite size effects start to appear. 

\begin{figure}[h!]
%\psfrag{x}[t][][1][0]{$g/J$}
%\psfrag{y}[b][][1][0]{$\bar{\mathcal{F}}$}
\includegraphics[width=7cm]{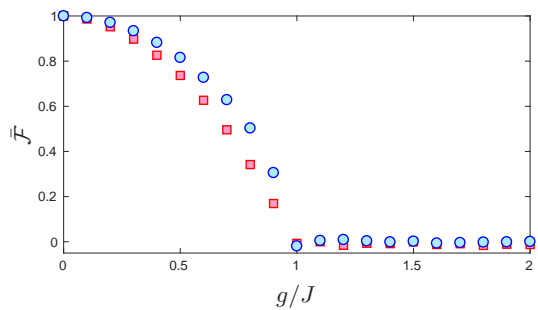}
\caption{The long time average of the order parameter OTO correlator is shown for the 
TFI chain in equilibrium from TEBD (blue circles) for $N=60$ and time window $60/J$ and after a quantum quench from a fully polarized state using ED (red squares) for $N=20$ and 
time window $20/J$, 
using numerical data similar to 
Eq. \eqref{timedepotoc} for several 
$g$'s.
\label{tfiotoc}}
\end{figure}

\paragraph{Steady state OTO correlator.}
As obvious from Fig. \ref{timedepotoc}, the steady state value of the OTO correlator can be determined accurately both from the TEBD and ED data by calculating the time 
average, $\bar{\mathcal{F}}$ as the $t\gg1/J$ limit
of $\frac 1 t\int_0^t \mathcal{F}(t')dt'$, albeit $t$ is still much smaller than the tunneling time (growing exponentially with $N$) 
between the almost degenerate ground states for finite $N$.
%Since from the numerics we can only access finite times, 
%we perform the time average over a finite time window.
The results for $\bar{\mathcal{F}}$ obtained in this way are shown in Fig. \ref{tfiotoc}.
We find that $\bar{\mathcal{F}}$ is nonzero in the ordered phase, and vanishes 
gradually upon approaching the equilibrium QPT, while it stays zero in the whole disordered, paramagnetic phase.
This happens not only in equilibrium, but also in the case of the quantum quench: the 
steady state value of the OTO correlator therefore serves as an a putative order parameter also for the DQPT.
Let us stress that this behaviour is in stark contract to the expectation value of 
$\langle\sigma_n^z(t)\rangle$ or $\langle\sigma_n^z(t)\sigma_n^z\rangle$, which both vanish for long times in the case of a quantum quench~\cite{essler}.
Importantly, one can detect the equilibrium QPT solely\cite{EPAPS} by performing a \emph{dynamical} measurement using 
OTO correlators without ever performing the challenging preparation of the actual ground state but rather 
doing a quantum quench from an initial condition that can be implemented with high fidelity in current experiments.

\paragraph{The ferromagnetic ANNNI model.} \jav{While the TFI is an integrable model, we now study a nonintegrable extension which is} the
transverse axial next-nearest-neighbor Ising (ANNNI) model, given by\cite{karrasch2013}
\begin{gather}
H=-J\sum_{n=1}^N\sigma^z_n\sigma^z_{n+1}-\Delta\sum_{n=1}^N\sigma^z_n\sigma^z_{n+2}+g\sum_{n=1}^N\sigma^x_n,
\label{tfinonint}
\end{gather}
where $\Delta$ denotes the strength of the second nearest neighbour interaction.
For $\Delta/J=0.5$, the Ising transition occurs at $g/J\approx 1.6$~\cite{karrasch2013}.
For $\Delta=0$, the model becomes integrable and reduces to Eq. \eqref{tfi}. For $J=0$, the model again reduces to two identical, independent copies of Eq. \eqref{tfi} for the even and odd sites.
For these two limiting cases, our previous results hold. For any finite $\Delta$, Eq. \eqref{tfinonint} becomes non-integrable\cite{alba,karrasch2013}.

\begin{figure}[h!]
%\psfrag{x}[t][][1][0]{$g/J$}
%\psfrag{y}[b][][1][0]{$\bar{\mathcal{F}}$}
%\psfrag{delta}[][][1][0]{$\Delta/J=0.5$}
%\psfrag{t1}[][][1][0]{equilibrium}
%\psfrag{t2}[][][1][0]{quench}
%\psfrag{N=12}[][][1][0]{$N=12$}
\includegraphics[width=8cm]{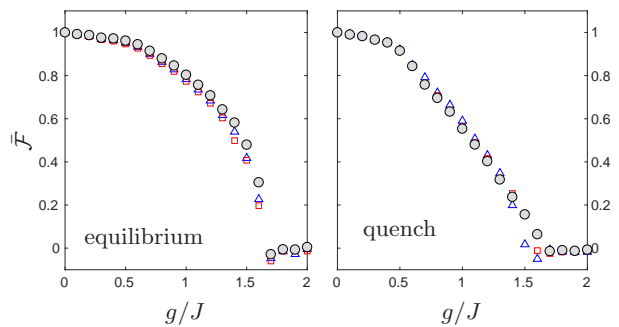}
\caption{The long time average of the order 
parameter OTO correlator is shown for the non-integrable 
TFI chain up until $tJ=20$ in equilibrium (left) for $N=20$, 30 and 40 (triangle, square and circle, respectively) and after a quantum quench from a fully polarized states (right) 
using ED for $N=12$, 16 and 20 
(triangle, square and circle, respectively), and $\Delta=0.5$. 
%TFI chain up until $tJ=20$ in equilibrium (blue circles) and after a quantum quench from a fully polarized states using ED (red squares) for $N=12$, 16 and 20 (small, medium 
%and large symbols, respectively), and $\Delta=0.5$. 
\label{tfiotocnonint}}
\end{figure}

We have calculated the OTO correlator of the order parameter using TEBD for the equilibrium case and using ED for the quantum quench. The results are plotted in Fig. \ref{tfiotocnonint}.
The OTO correlator behaves similarly to the integrable case: the imaginary part vanishes for long times in equilibrium and is identically zero after a quench, thus
we focus only on its real part $\mathcal{F}_R$. This  takes a finite value in the ferromagnetic phase both in equilibrium or after the quench, and vanishes on the paramagnetic side.
Therefore, the identification of the OTO correlator as a putative order parameter works ideally for non-integrable systems as well.

\paragraph{The fully connected  transverse field Ising chain: the Lipkin-Meshkov Glick model.}
Finally, we turn to the Lipkin-Meshkov Glick model \cite{Lipkin}, describing the fully connected version of Eq. \eqref{tfi}.
The Hamiltonian of this system can
be expressed in terms of the collective spin operators $S^\alpha=\sum_{i=1}^N\sigma_i^\alpha/2$, $\alpha=x,y,z$, as 
\begin{gather}
H_{LMG}=-\frac JN \left(S^z\right)^2+gS^x \, .
\end{gather}
This model exhibits not only a quantum phase transition in the ground state at $g/J=1$ but also a symmetry-broken phase and respective transition at nonzero temperatures.
\jav{Consequently, this system allows us to study the  dynamics of the OTO correlator in the presence of symmetry-breaking at excited energy densities above the ground state, which is absent for the short-range model discussed before and
which leads also to a DQPT} at $g/J=1/2$ for quantum quenches when initializing the system in the fully polarized state~\cite{Sciolla2011,Zunkovic2016,Zhang20172}.
This DQPT separates a regime of nonzero value of the order parameter $S^z$ in the steady state for $g/J<1/2$ from a disordered phase for $g/J>1/2$ where the order parameter vanishes.
Importantly, the dynamics of the LMG Hamiltonian can be realized in systems of trapped ions~\cite{Lanyon2011, Britton2012, Richerme2014, Jurcevic2014, Jurcevic2017}.
The LMG model is exactly solvable since $[\vec S^2,H_{LMG}]=0$ such that the Hamiltonian decomposes into disconnected blocks for each of the total spin quantum number $S$.
Due to its exact solvability the system is integrable and not thermalizing.
As a consequence the anticipated DQPT after a quantum quench is not thermal but rather a genuine nonequilibrium transition without equilibrium counterpart.

\begin{figure}[h!]
%\psfrag{x}[t][][1][0]{$Jt/S$}
%\psfrag{y}[b][][1][0]{$\mathcal{F}_R(t)$}
%\psfrag{t1}[t][][1][0]{equilibrium}
%\psfrag{t2}[t][][1][0]{quench}
\includegraphics[width=8cm]{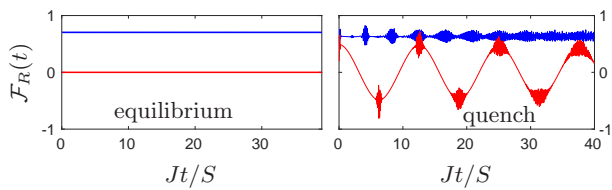}
\caption{The representative time evolution of the real part of OTO correlator is
shown for the LMG model in equilibrium (left panel) and after a quench (right panel) from the fully polarized state in the ordered phase with $g/J=0.4$ (blue) and
disordered phase with  $g/J$=1.2 (red) for $N=499$.
\label{lmgotoctime}}
\end{figure}

\begin{figure}[h!]
%\psfrag{x}[t][][1][0]{$g/J$}
%\psfrag{y}[b][][1][0]{$\bar{\mathcal{F}}$}
\includegraphics[width=7cm]{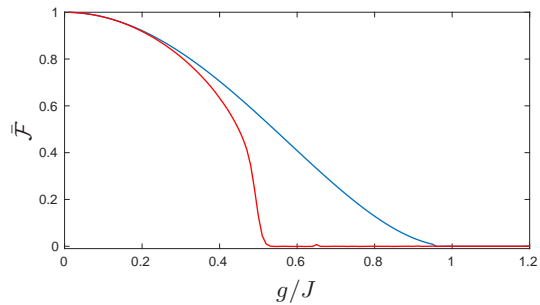}
\caption{The long time average of the order parameter OTO correlator is shown for the LMG model with $N=1599$ in equilibrium (blue) 
and after a quantum quench (red) from ED within the time window $20S/J$.
\label{lmgotoc}}
\end{figure}

In the following, we consider the maximum spin sector $S=N/2$ which also contains the fully polarized initial condition we consider for the quantum quench.  
As already mentioned in Eq. \eqref{defM}, we calculate the OTO correlator in the LMG model for $\mathcal{M} = S^z/S$.
A typical time evolution is depicted in Fig. \ref{lmgotoctime}, while the time averaged value\footnote{Here, time average is taken for $Jt\gg S$}
of the OTO correlator is shown in Fig. \ref{lmgotoc}.
From the data one can clearly see that the OTO correlator can both detect the equilibrium as well as dynamical transition.
Compared to the previously discussed models there is, however, an apparent difference.
\jav{While the equilibrium $\mathcal{F}$ still diagnoses the QPT, the $\mathcal{F}$ after a quantum quench signals the DQPT
 suggesting that the detection of the ground state phase transition from quantum dynamics is limited to the short-range models discussed before.}
Consequently, we find that the nature of the critical point probed by the OTO correlator depends on the initial condition.
Whether it is possible to also detect the thermal transition remains an open question.

\paragraph{Concluding discussion.}
In this work we have shown \jav{numerical evidence that} OTO correlators can be used to dynamically detect both 
equilibrium as well as dynamical quantum critical points in one-dimensional short-range and infinite-range transverse-field Ising models.

\jav{In addition of serving as an order parameter, our results for the OTO correlation have \jav{further} ramifications as well. For the operators considered for the TFI and ANNNI model, 
$C(t)=2(1-\textmd{Re}\mathcal{F}(t))$ holds.
In Refs. \cite{roberts,roberts2016}, it was argued that in a suitably chaotic system, the OTO correlator is expected to vanish and the OTO commutator to approach
$C(t\rightarrow\infty)\approx 2\langle V^2\rangle\langle W^2\rangle$ for any nonzero temperature state.
This is exactly what we find in the \emph{disordered} phase of both models for the ground state as well for a quantum quench, as shown in Figs. \ref{tfiotoc} and \ref{tfiotocnonint}:
the steady state value of the OTO correlator vanishes for $g>J$, therefore $C(t\rightarrow\infty)\rightarrow 2$ in the exact same manner as is expected in chaotic systems.
In the ordered phase the situation is, however, different.
Both the ground-state and the quantum quench OTO correlators are nonzero.
For the considered model, there are two possible explanations for the apparent discrepancy to the conjectured generic long-time dynamics.
First, the arguments hold for a generic operator, but the order parameter might not fall under this category.
Second, our conclusions hold as long as our time averaging scheme over a finite time window is suitable and there appears no fundamental change of the OTO correlator 
dynamics at larger times, for which we also do not find evidence.
In this worst-case scenario, our observations would hold over an extended metastable state on intermediate time scales.
Summarizing, we find from our study that the OTO correlator dynamics can show unexpected behavior whenever the system exhibits a symmetry-broken phase which is 
precisely at the heart of making it a tool to dynamically detect quantum phases. % \jav{in transverse-field Ising chains}.
}

\jav{
For the future it is an interesting question of how the OTO correlator for the order parameter behaves for other systems than the studied Ising chains. 
Moreover, it is currently unclear what would happen for the OTO correlator for transverse-field Ising models in two and three spatial dimensions, which 
exhibit symmetry-broken phases at nonzero temperature and in addition are also expected to thermalize in the long-time limit, as opposed to the long-range 
Ising model studied in this work.
Specifically, the OTO correlator either detects the ground state or the thermal transition for which we cannot make a conclusive prediction within our numerics~\cite{EPAPS}.
}

In the remainder we now discuss how these observations can be accessed in quantum simulators experimentally.
While the considered quantum quench dynamics of the short-range Ising models can be synthesized in Rydberg atoms\cite{Bernien2017,Guardado2017}, the long-range version can be realized in trapped ions~\cite{Jurcevic2014,Jurcevic2017,Richerme2014, Smith2016, Zhang2017, Zhang20172}.
%
%Compared to the measurement of conventional observables, the OTO correlators, however, require an adapted experimental sequence.
%
The fully-polarized initial condition $|\psi_0\rangle = |\uparrow \uparrow\uparrow\dots\rangle$ can be realized with high fidelity~\cite{Lanyon2011,Schauss2012,Jurcevic2017,Bernien2017,Guardado2017, Zhang2017, Zhang20172}.
Since $|\psi_0\rangle$ is an eigenstate of $\mathcal{M}$, see Eq.~(\ref{defM}), we only have to consider a reduced quantity $\tilde{\mathcal{F}}(t) = \langle \psi_0 | \mathcal{M}(t) \mathcal{M} \mathcal{M}(t) | \psi_0\rangle$.
For $\mathcal{M}=\sigma_n^z$ we can reexpress $\tilde{\mathcal{F}}(t) = \langle \psi_t |  \mathcal{M}  | \psi_t\rangle$ as a conventional expectation value with $ |\psi_t\rangle = U^\dag(t) \exp[i \frac{\pi}{2}\sigma_n^z] U(t) |\psi_0\rangle$ where $U(t) = \exp[-iHt]$ using $\exp[i \frac{\pi}{2}\sigma_n^z] = i\sigma_n^z$, see also~\cite{garttner}.
Consequently, it would be necessary to apply a sequence of unitary transformations implementing i) a time evolution with the Hamiltonian $H$, ii) a local single qubit rotation on spin $n$, and iii) a backward time evolution with Hamiltonian $-H$.
%
%All the requirements for this sequence have been separately realized such that the protocol appears feasible within current experimental technology.
%
An additional challenge is that the total simulation time is doubled to $2\times t$ due to forward an backward evolution.
Fortunately, clear signatures of the order
parameter can be estimated already from the data on times $t \lesssim 5J^{-1}$, see Fig.~\ref{timedepotoc}, which is close to the accessible experimental range~\cite{Guardado2017,Bernien2017,Zhang20172}.

\begin{acknowledgments}

This research is supported by the National Research, Development and Innovation Office - NKFIH within the Quantum Technology National Excellence Program (Project No.
      2017-1.2.1-NKP-2017-00001),  K105149, K108676, SNN118028 and K119442
and by Romanian UEFISCDI, project number PN-III-P4-ID-PCE-2016-0032.
FP acknowledges the support of the DFG Research Unit FOR 1807 through grants no. PO 1370/2- 1, TRR80, and the Nanosystems Initiative Munich (NIM) by the German Excellence Initiative.
M. H.  acknowledges  support  by  the  Deutsche Forschungsgemeinschaft  via  the  Gottfried
Wilhelm Leibniz Prize program.

\end{acknowledgments}

\bibliographystyle{apsrev}
\bibliography{literature}

\setcounter{equation}{0}
\renewcommand{\theequation}{S\arabic{equation}}
\setcounter{figure}{0}
\renewcommand{\thefigure}{S\arabic{figure}}

\section{Supplementary material for "Detecting equilibrium and dynamical quantum phase transitions in Ising chains via out-of-time-ordered correlators"}

\section{The OTO correlator signals the equilibrium QPT after a quench for short range interacting models}

\begin{figure}[h!]
\psfrag{x}[t][][1][0]{$t$}
\psfrag{y}[b][][1][0]{{\color{red}$\mathcal{F}_R(t)$}, {\color{blue}$2r(t)$}}
\includegraphics[width=7cm]{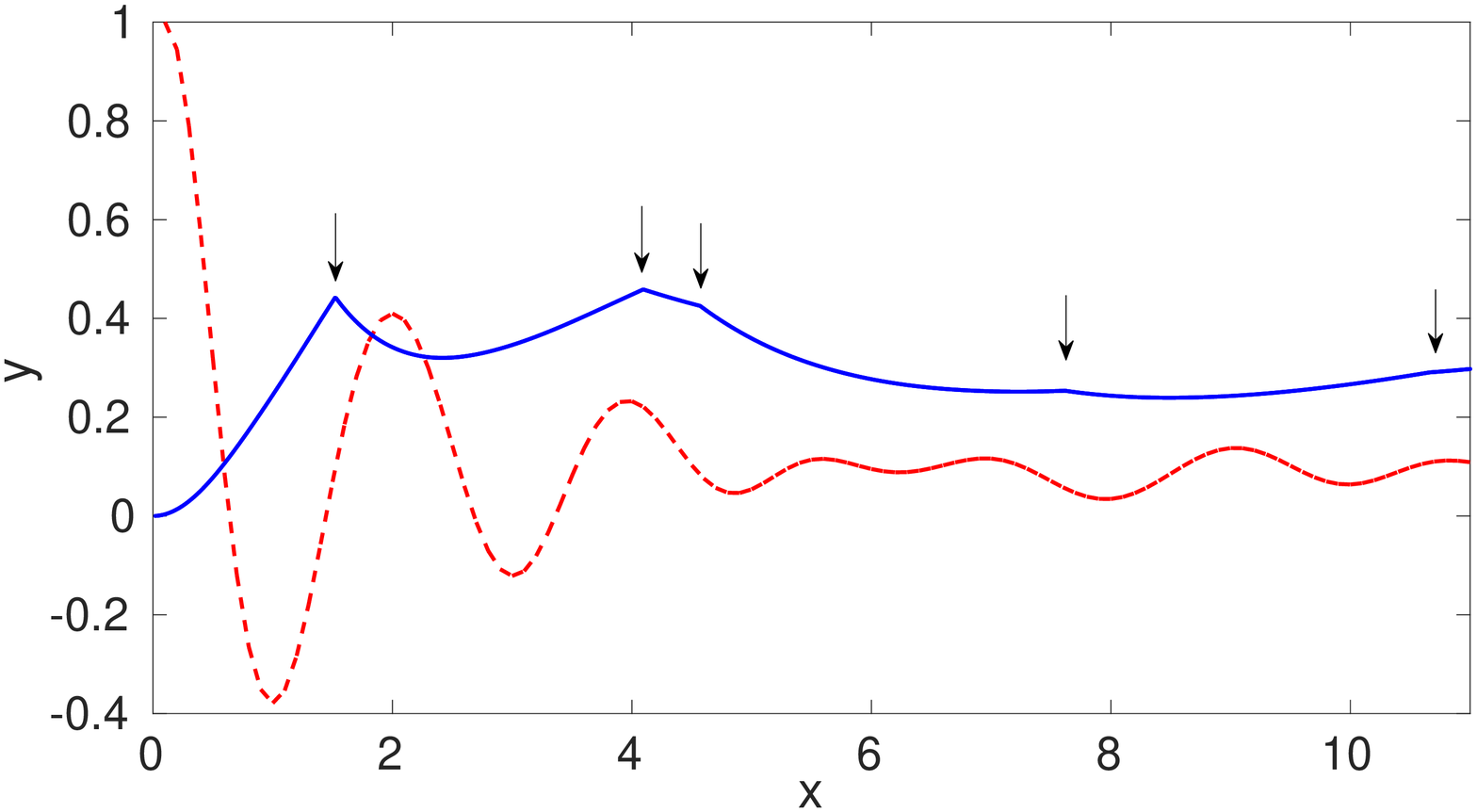}
\caption{The OTO correlator (red dashed line) from ED for $N=20$ and twice the rate function (blue solid line) from Ref. \onlinecite{vajna2014}
for the XY model are plotted for a quench within the ordered phase with  $(h_0,\gamma_0)=(0,0.3)$ and $(h_1,\gamma_1)=(0.6,0.2)$.
While the rate function signals a DQPT (the non-analytic points are denoted by arrows from $\partial r(t)/\partial t$), the OTO correlator stays also finite at late times.
\label{loschmidtotocxy}}
\end{figure}

For many short range interacting models, such as the TFI and ANNNI models, studied in the main text, the equilibrium and dynamical quantum phase transition (DQPT) points coincide. Therefore, it is a natural question to ask, which one of these
is indeed signaled by the OTO correlator? 
To answer this question we turn to the 1D XY model in a magnetic field, which has been studied in Ref. \onlinecite{vajna2014}, and was demonstrated that for certain parameter range DQPT can occur without crossing an equilibrium
phase boundary, thus without an equilibrium QPT counterpart.
Its Hamiltonian is
\begin{gather}
H_{XY}=\sum\limits_{j=1}^{N-1} \frac{1+\gamma}{2}\sigma_j^x\sigma_{j+1}^x+ \frac{1-\gamma}{2}\sigma_j^y\sigma_{j+1}^y-h\sum\limits_{j=1}^{N}\sigma_j^z.
\end{gather}
For $\gamma>0$ and $h<1$, the spins are ordered in the $x$ direction and quenching
from an initial value $(h_0,\gamma_0)$ to a final $(h_1,\gamma_1)$ within the same equilibrium phase, DQPT show up for
\begin{gather}
2\gamma_0\gamma_1<1-h_0h_1-\sqrt{(h_0^2-1)(h_1^2-1)},
\end{gather}
as was calculated in Ref. \onlinecite{vajna2014}. Thus, by choosing $h_0=0$, $\gamma_0=0.3$ and e.g. $\gamma_1=0.2$, a DQPT occurs for $h_1>0.475$, though the equilibrum phase boundary is located at $h_1=1$.
Following Ref. \onlinecite{vajna2014}, we have calculated the Loschmidt overlap, i.e.
\begin{gather}
G(t)=\langle\Psi(h_0,\gamma_0)|\exp(itH(h_1,\gamma_1))|\Psi(h_0,\gamma_0)\rangle
\end{gather}
for $(h_0,\gamma_0)=(0,0.3)$ and $(h_1,\gamma_1)=(0.6,0.2)$. The rate function, $r(t)=-\lim_{N\rightarrow\infty} \frac{1}{N}\ln|G(t)|^2$ is shown in Fig.  \ref{loschmidtotocxy}
together with the OTO correlator of the order parameter, $\sigma^x_n$ using ED for N=20.

While cusps show up in the rate function, indicating the occurence of DQPT, the OTO correlator of the order parameter stays finite at late times, and only vanishes upon crossing the QPT, similarly to the TFI models.

\section{Order parameter correlation function} 

We address the behaviour of the correlation function of the order parameter,
\begin{gather}
\chi(t)=\langle \mathcal{M}(t)\mathcal{M}\rangle
\end{gather}
for several variants of the transverse field Ising model both in equilibrium and after a quench from the fully polarized state.
Note that for the latter case, the wavefunction is an eigenstate of the order parameter as $\mathcal{M}|\uparrow \uparrow\uparrow\dots\rangle=|\uparrow \uparrow\uparrow\dots\rangle$,
thus the above correlation function measures simply the decay of the initial magnetization, i.e. 
\begin{gather}
\chi(t)= \langle\dots\uparrow \uparrow\uparrow|\mathcal{M}(t)|\uparrow \uparrow\uparrow\dots\rangle \textmd{  after a quench.}
\end{gather}

In equilibrium, for large temporal separation, $\chi(t)$ measures the square of the magnetization and is finite/zero in the ordered/disordered phases, respectively\cite{sachdev}. 
In contrast, after a quench, it vanishes with time\cite{essler} since the quench heats up the system and no long range order is possible for 1D short range interacting models at finite temperatures, as we demonstrate below.

\begin{figure}[t!]
\psfrag{x}[t][][1][0]{$tJ$}
\psfrag{y}[b][][1][0]{Re$\chi(t)$}
\psfrag{t1}[][][1][0]{equilibrium}
\psfrag{t2}[][][1][0]{quench}
\psfrag{t3}[][][1][0]{equilibrium}
\psfrag{t4}[][][1][0]{quench}
\includegraphics[width=8cm]{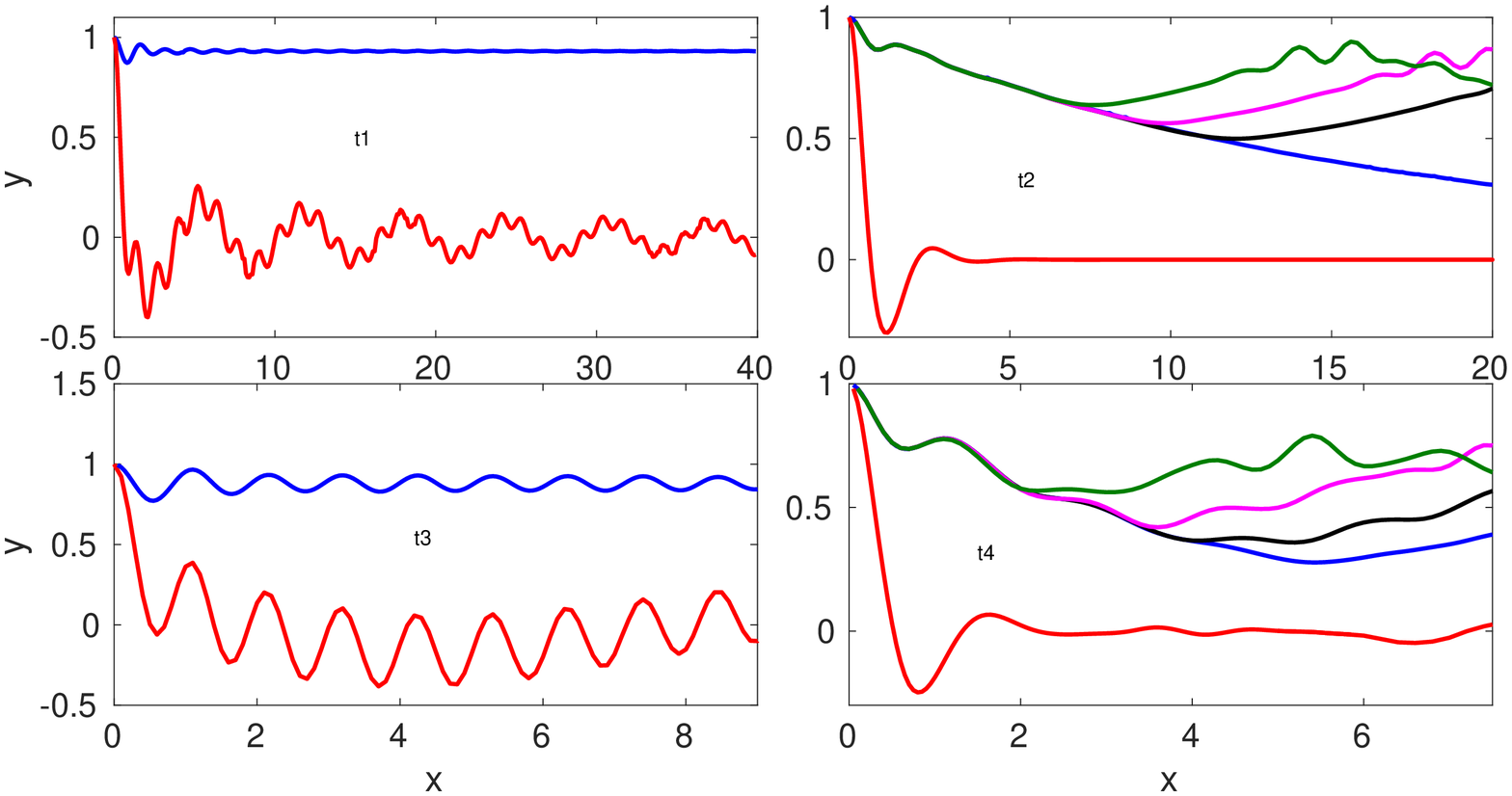}
\caption{The correlation function of the order parameter in equilibrium and after a quench is shown.
The top panels depict the TFI with $N=60$ from TEBD for $g/J=0.5$ (blue) and 1.5 (red).
For comparison, the ED data with $N=22$ (black), 18 (magenta) and 14 (green) is also shown after a quench within the ordered phase.
The bottom panels visualize the response of the ANNNI model from ED with $g/J=1$ (blue) and 2 (red) with $N=22$.
The  ED data with $N=18$ (black), 14 (magenta) and 10 (green) is also shown after a quench within the ordered phase, paralleling closely to the behaviour of the TFI model.
\label{tfimtm0}}
\end{figure}

Let us start with the transverse field Ising model, Eq. (4) in the main text. 
The $\chi(t)$ correlation function in equilibrium signals the ordering, reaching a finite steady state value in the ordered phase while vanishing in the disordered phase, as shown in Fig. \ref{tfimtm0}.
After a quench, however, the very same correlation function, measuring now $\langle \mathcal{M}(t)\rangle$ in the polarized state, vanishes identically in the long time limit. In particular, the
numerical data for $g/J=0.5$ after the quench is well fitted by $0.95\exp(-0.056 Jt)$, in agreement with the expected exponential decay of the magnetization.
The finite size data also follows this trend before the finite size effects kick in.

The ANNNI model produces qualitatively similar behaviour: while the steady state value of the equilibrium $\chi(t)$ indicates the ordered/disordered phase,
it vanishes after the quench, therefore its steady state value cannot indicate the equilibrium quantum critical point and cannot serve as a putative order parameter for the DQPT.
As we have demonstrated in the main text, the OTO correlator, on the other hand, fulfills this job perfectly.

Finally, for the sake of completeness,
 we also show  $\chi(t)$ for the LMG model. This model has a finite temperature phase transition, thus the correlation function of the order parameter detects not only the equilibrium but also the DQPT, in which case it measures
directly the magnetization. In this context, one does not gain much by studying the OTO correlator as simpler correlators contain already information about ordering, nevertheless to treat long range interacting models
on equal footing as short range models, it is important to emphasize that the OTO correlator serves as a universal diagnostic tool for equilibrium and DQPT, unlike $\chi(t)$.

\begin{figure}[h!]
\psfrag{x}[t][][1][0]{$tJ/S$}
\psfrag{y}[b][][1][0]{Re$\chi(t)$}
\psfrag{t1}[][][1][0]{equilibrium}
\psfrag{t2}[][][1][0]{quench}
\psfrag{t3}[][][1][0]{equilibrium}
\psfrag{t4}[][][1][0]{quench}
\includegraphics[width=8cm]{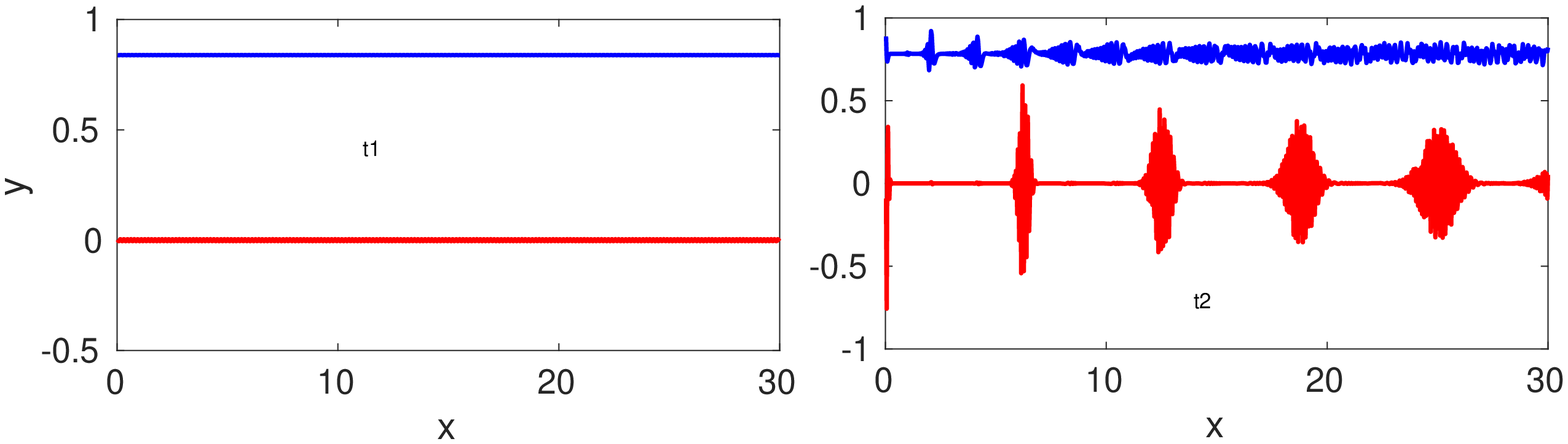}
\caption{The representative time evolution of the correlation function of the order parameter is
shown for the LMG model in equilibrium (left panel) and after a quench (right panel) from the fully polarized state in the ordered phase with $g/J=0.4$ (blue) and
disordered phase with  $g/J$=1.2 (red) for $N=499$.
\label{lmgmtm0}}
\end{figure}

\begin{figure}[b!]
\psfrag{x}[t][][1][0]{$tJ$}
\psfrag{y}[b][][1][0]{Re$\mathcal{F}(t)$}
\includegraphics[width=8cm]{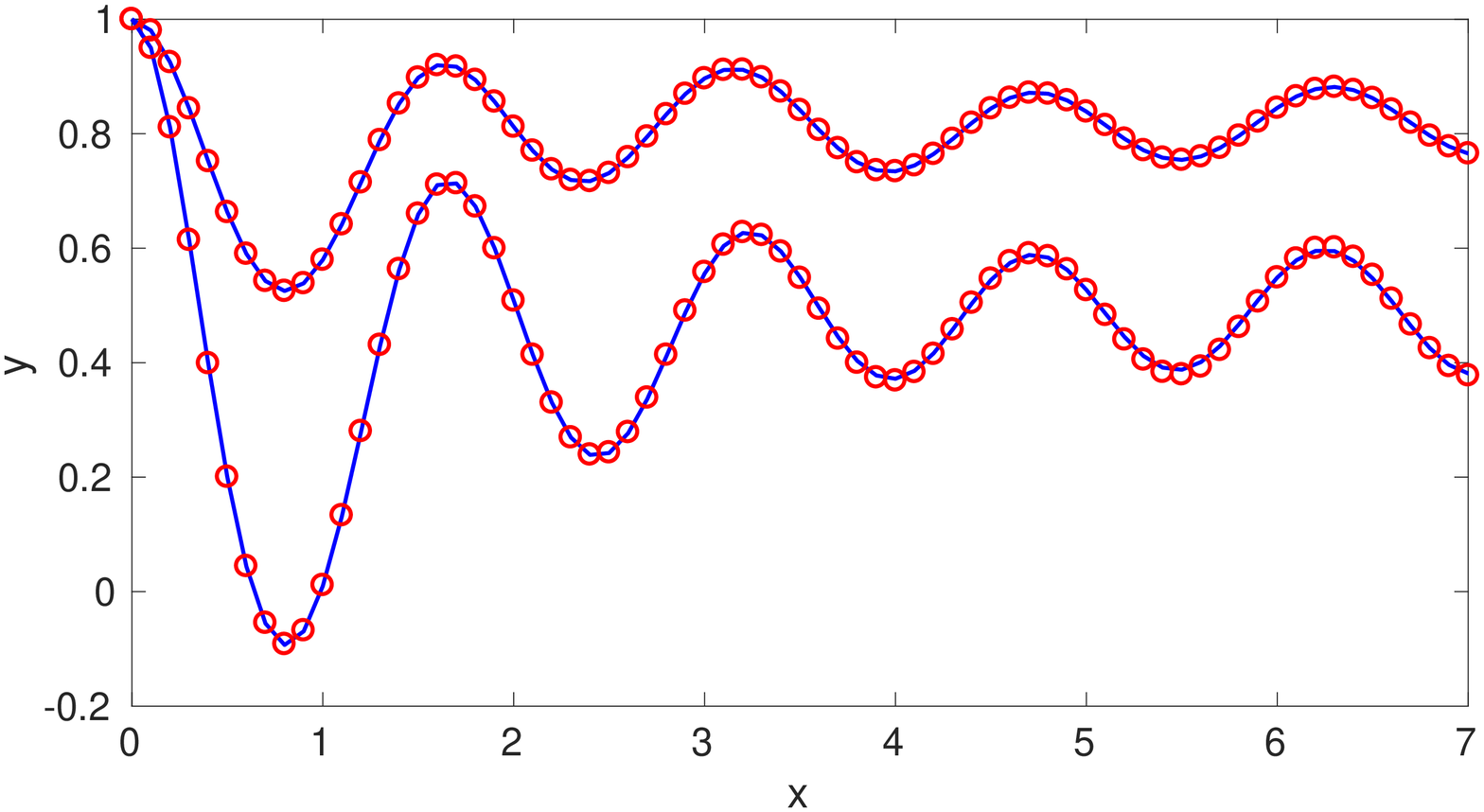}
\caption{The real part of OTO correlators for TFI with $N=14$ from ED are
shown in equilibrium in the ordered phase with $g/J=0.5$ (top) and 0.8 (bottom) using an unpolarized ground state with $J'=0$ (blue solid line)
and a polarized ground state with $J'=J/100$ (red circles). The respective homogeneous
magnetizations per site are 0.965 and 0.786 for $J'=J/100$ and zero for $J'=0$. In spite of the completely different magnetizations, the OTO correlators are identical.
\label{tfipolarizedvsunpolarized}}
\end{figure}

\begin{figure}[b!]
\psfrag{x}[t][][1][0]{$tJ$}
\psfrag{y}[b][][1][0]{{\color{blue}$\langle\mathcal{M}(t)\rangle$}, {\color{red}Re$\mathcal{F}(t)$}}
\includegraphics[width=8cm]{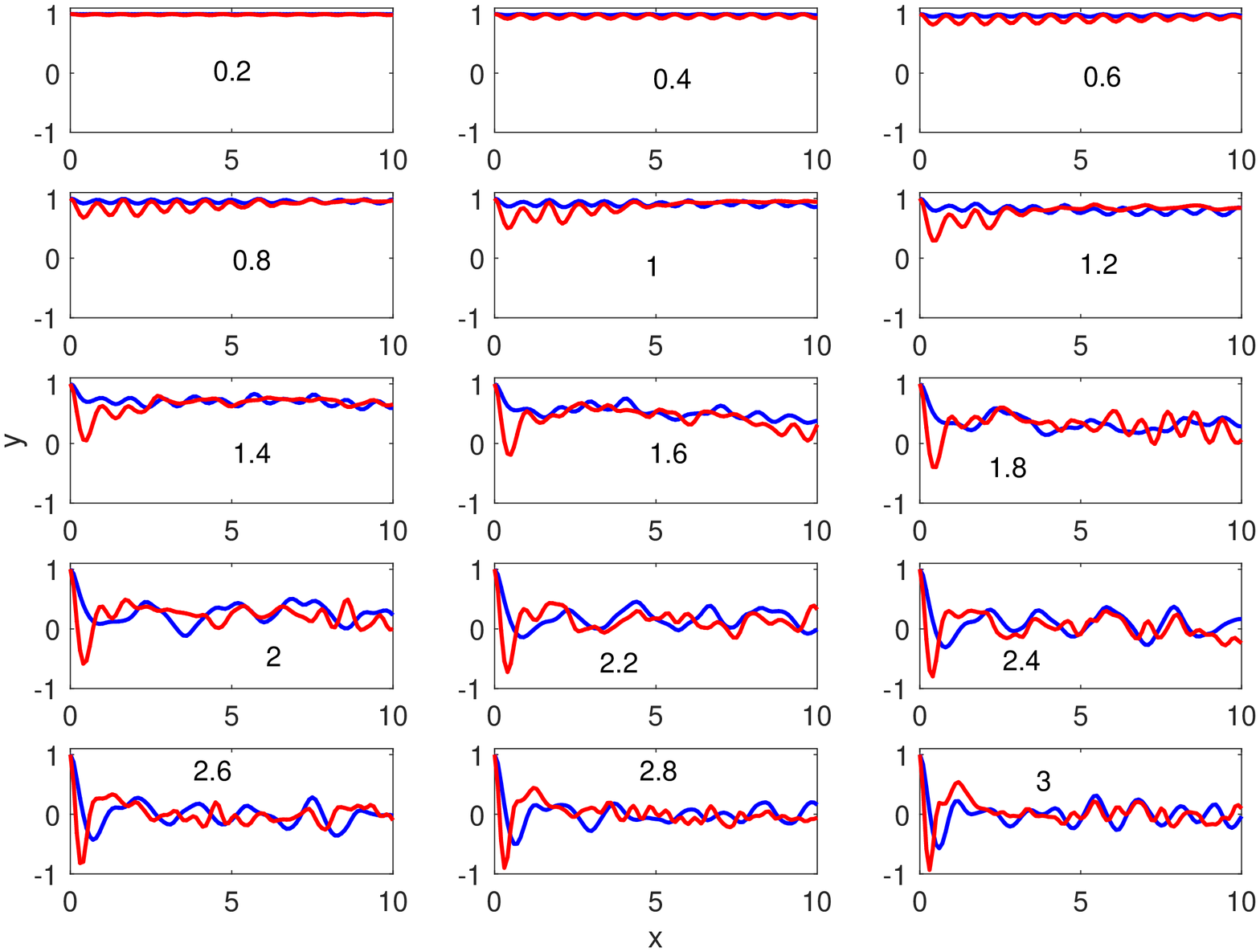}
\caption{The magnetization (blue) and the real part of OTO correlator (red) for 2D TFI on $4\times 4$ square lattice from ED are
shown after quench from a polarized state. The $g/J$ ratio of the final Hamiltonian is indicated in each panel. }
\label{2dmotoc}
\end{figure}

\section{Polarized vs. unpolarized state}

By numerically diagonalizing the transverse field Ising model in Eq. (4) using periodic boundary conditions, the ground state wavefunction is a linear superposition of the two symmetry broken ground states in the ordered phase, such that
the magnetization vanishes for this wavefunction. By adding a small source field to a given site for the transverse field Ising model of the form  $H'=-J'\sigma^z_1$ with $J'\ll J,g$, this additional term facilitates symmetry breaking
and the expectation value of the magnetization using this wavefunction gives the actual order parameter.
On the other hand, a tiny source field does not yield any magnetization in the disordered phase.
The OTO correlator is insensitive whether we take the polarized ground state wavefunction or the superposition with the symmetry broken ground states with no magnetization, as shown in Fig. \ref{tfipolarizedvsunpolarized}.

\section{2D TFI model}

In this section, we elaborate  on whether the OTO correlator signals the ground state or thermal phase transition.
To this end, we investigate the two dimensional TFI model, which possesses a thermal as well as quantum phase transition and is
non-integrable, thus it is expected to thermalize after a quench.
Its  magnetization after a quench from the polarized state indicates the dynamical phase transition, similarly to the LMG model in the main text, 
what we compare to the behaviour of the OTO correlator.
Its Hamiltonian with periodic boundary condition is given by
\begin{gather}
H=-J\sum_{\langle R,R'\rangle}\sigma^z_R\sigma^z_{R'}+g\sum_{R}\sigma^x_R,
\label{2dtfi}
\end{gather}
where $R$ and $R'$ are lattice vectors for the 2D square lattice and $\langle R,R'\rangle$ denotes nearest neighbour pairs  such, that each pair is condidered only once.

The model in Eq. \eqref{2dtfi}
exhibits an equilibrium quantum phase transition\cite{croo,blote} at $g/J\approx 3.04$ from a ferromagnetic state to a paramagnetic phase with increasing $g$.
We focus on the time dependence of the magnetization, $\mathcal{M}=\sigma^z_R$, after a quench from a fully polarized state, as well as the corresponding OTO correlator for a
$4\times 4$ square lattice using ED.
The results are plotted in Fig. \ref{2dmotoc}. While the precise location of the DQPT cannot be determined unambiguously due to the small lattice size, 
it seems that
the magnetization and the OTO correlator follow the same behaviour, thus both signal the DQPT transition equally well. The DQPT is located at around $g/J=2-2.5$, and for small 
$g$, both the magnetization and the OTO correlator saturates to a finite, non-zero value, while they oscillate around zero for $g/J>2.5$.
This indicates that the OTO correlator in models with broken symmetry states at finite temperature signals the thermal and not the ground state phase diagram, though
our results are far from being conclusive due to the small lattice size we consider.

\end{document}